\documentclass[prb]{revtex4}

\usepackage{graphicx}
\usepackage{dcolumn}
\usepackage{bm}

\begin{document}


\title{Donor Electron Wave Functions for Phosphorus in Silicon: Beyond Effective Mass Theory}

\author{C.J. Wellard and L.C.L. Hollenberg}

\affiliation{Centre for Quantum Computer Technology,\\
School of Physics, University of Melbourne, Victoria 3010,
AUSTRALIA.}

\date{\today}

\begin{abstract}
We calculate the electronic wave-function for a phosphorus donor in silicon by numerical diagonalisation of the donor Hamiltonian in the basis of the pure crystal Bloch functions. The Hamiltonian is calculated at discrete points localised around the conduction band minima in the reciprocal lattice space. Such a technique goes beyond the approximations inherent in the effective-mass theory, and can be modified to include the effects of altered donor impurity potentials, externally applied electro-static potentials, as well as the effects of lattice strain. Modification of the donor impurity potential allows the experimentally known low-lying energy spectrum to be reproduced with good agreement, as well as the calculation of the donor wavefunction, which can then be used to calculate parameters important to quantum computing applications.

\end{abstract}

\pacs{Valid PACS appear here}
\maketitle

\section{\label{sec:introduction} Introduction}
The electronic states of phosphorus donors in silicon are of increasing interest in the field
of quantum computing due to the fundamental role they play in several promising proposals for
a scalable quantum information processor. The precise nature of the
qubits differ between proposals, and include nuclear spin\cite{Kane98},
electron spin\cite{Vrijen00}, or the low-lying electronic states of an
ionised two-donor system\cite{Hollenberg04}. Common to all these
proposals however, is the way in which quantum information processing is
implemented, through external coherent control of the electron
wave-function of the phosphorus donor. The rigours of implementing
large-scale quantum algorithms on such a device are such that these
wave-functions must be controlled to a remarkable precision - for example,
standard estimates require an error of no more than one part in ten
thousand per fundamental quantum operation. To meet these stringent requirements it is vital to have a detailed
understanding of the electronic wave-function of the phosphorus
donor. Although the study of these states is quite mature, this level of understanding is currently lacking.

Experimentally, the spectrum of these donor electron states is quite
well known \cite{Morin53,Picus56,Aggarwal65}, and much sophisticated
theory has been developed to try and understand these results. The standard 
description of these donor states, the effective-mass theory of Kohn and Luttinger
\cite{Kohn57,Kohn55,Luttinger55}, has been very successful in predicting the qualitative nature
of these states, however, precise numerical agreement with experiment has been elusive. The theory is based around the expansion of the wave-function in the Bloch states of the pure silicon crystal. The coefficients of this expansion form an envelope function, and it is the reduction of the equation determining this envelope function to a set of six, non-isotropic, hydrogen-like equations that forms the core of the effective-mass approach. Modifications to this model can then be made to include effects such as the valley-orbit coupling that is responsible for the lifting of the six-fold ground-state degeneracy observed experimentally. Generally however, these additions to the theory are handled in a somewhat {\it ad-hoc} manner, and are often inconsistent with the approximations made in deriving the theory in the first place.

In this article we address these issues by applying a numerical approach where some of the simplifications inherent in the effective-mass theory are avoided, and the exact Hamiltonian describing the donor electron is solved, using a truncated basis. This approach allows the inclusion of non-Coulombic donor potentials, which can provide the valley-orbit coupling necessary to lift the ground-state degeneracy, in a manner that is consistent with the theory. 

The paper is organised as follows: We begin with a review of the standard effective-mass theory in section \ref{sec:eff_mass}, paying particular attention to the approximations inherent in the treatment. In section \ref{sec:bloch} we outline our numerical approach to the problem, focusing on the construction of the donor Hamiltonian matrix, and the choice of Bloch-basis states. Section \ref{sec:correction} contains results for a corrected impurity potential, which provides the necessary valley-orbit coupling to reproduce the known ground-state energy, and approximate the energy of the low-lying excited states. In section\ref{sec:exchange} the wave-function obtained from these calculations is used to calculate the strength of the exchange coupling between neighbouring donors, as a function of the donor separation. These results are compared with results obtained using Kohn-Luttinger type effective-mass wave-functions for a Coulombic impurity potential \cite{Koiller02A,Koiller02B,Wellard03}. The effect of the non-Coulombic impurity potential is to increase the  localisation of the donor wave-function, which leads to a reduced exchange coupling when compared to estimates based on a Coulombic impurity potential. In section \ref{sec:strain} we calculate the electronic wave-function for donors in
uni-axially stressed silicon, and use the resulting wave-function to
calculate the exchange coupling between neighbouring donors. We
qualitatively reproduce the results of Koiller {\it et al}
\cite{Koiller02B} which show that this strain can be used to
eliminate the exchange oscillations for donors in the same [001]
plane, and that the strength of the interaction is generally increased when compared to the unstrained case. Section \ref{sec:epot} illustrates how this numerical  method also has the freedom of including externally applied electric potentials, such as those generated by the control gates used to manipulate the state of the donor electron with the goal of implementing quantum information processing, in a straight-forward fashion. We solve for the case
of a uniform applied electric field, and calculate the shift in the
electron-nuclear contact hyperfine coupling strength, as both a
function of the strength of the applied field, and the distance of
the the donor from a silicon-oxide barrier. We compare these results with those obtained using 
a tight-binding approach \cite{Martins04} in which the donor wave-function is solved by expanding in a basis of states localised in real-space, and extending over several hundreds of thousands of silicon atoms. We finish with some concluding remarks and a brief discussion of some potential applications for this approach.

\section{\label{sec:eff_mass} Effective Mass Theory}
The theory of shallow donor electron states in silicon was developed by Kohn
and Luttinger in the  mid 1950's \cite{Luttinger55,Kohn57,Fritzsche62}, in what
has become known as effective mass theory. In this section we will briefly
review this theory, a more detailed discussion can be found in \cite{Pantelides78}.
\par The aim is to find the low energy solutions of the Schr\"odinger equation for the donor electron
 wave-function;
\begin{equation}
(H_0 + U({\bf r}))\psi({\bf r}) = E\psi({\bf r}),
\label{equation:schroedinger1}
\end{equation}
where $H_0$ is the Hamiltonian for the pure silicon crystal and
$U({\bf r})$ is the impurity potential due to the presence of the
phosphorus ion. To this end it is natural to expand solutions  in terms of the eigenstates of the pure crystal Hamiltonian, 
the Bloch functions for silicon: $\phi_{\bf k}({\bf r})$,
\begin{equation}
\psi({\bf r}) = \int F({\bf k}) \phi_{\bf k}({\bf r}) d{\bf k}.
\end{equation}
Substituting this into Eq~\ref{equation:schroedinger1}, multiplying from the
left by $\phi^*_{{\bf k}'}({\bf r})$ and integrating over all space leads
to
\begin{equation}
E F({\bf k}') = E_{{\bf k}'} F({\bf k}') + \int \phi^*_{{\bf
k}'}({\bf r}) U({\bf r}) \phi_{\bf k}({\bf r}) F({\bf k}) d{\bf k}
d{\bf r}, \label{equation:schroedinger2}
\end{equation}
where we have used the eigen-structure of the pure crystal,
$H_0\phi_{{\bf k}'}({\bf r})= E_{{\bf k}'} \phi_{{\bf k}'}({\bf
r})$. The Bloch states can be
expanded in terms of functions that share the periodicity of the
crystal lattice $\phi_{\bf k}({\bf r}) = {\rm e}^{i {\bf k}.{\bf r}}
u_{\bf k}({\bf r})$, with $u_{\bf k}({\bf r}) = \sum_{\bf G} A_{{\bf
k},{\bf G}} {\rm e}^{i {\bf G}.{\bf r}}$. The ${\bf G}$ are
reciprocal lattice vectors of the silicon crystal, and have the
property ${\rm e}^{i {\bf G}.{\bf R}} = 1$, where ${\bf R}$ is an
integer multiple of the fundamental translation vectors of the
crystal. The coefficients $A_{{\bf k},{\bf G}}$ are obtained using a pseudo-potential method which has been parametrised to reproduce the band-structure of the pure silicon crystal \cite{Cardona01}. Expanding the Bloch functions in
 Eq~\ref{equation:schroedinger2} yields
\begin{eqnarray}
E F({\bf k}') &=& E_{{\bf k}'} F({\bf k}') + \int \sum_{{\bf G},{\bf G}'}
A^*_{{\bf k}',{\bf G}'} A_{{\bf k},{\bf G}}
{\rm e}^{i ({\bf k}-{\bf k}').{\bf r}} {\rm e}^{i ({\bf G}-{\bf G}').{\bf r}}
U({\bf r})  F({\bf k}) d{\bf k} d{\bf r}  \nonumber\\
&=& E_{{\bf k}'} F({\bf k}') + \int \sum_{{\bf G},{\bf G}'}
A^*_{{\bf k}',{\bf G}'} A_{{\bf k},{\bf G}}
{\tilde U}({\bf k}-{\bf k}'+{\bf G}-{\bf G}')  F({\bf k}) d{\bf k},
\label{equation:schroedinger_exact}
\end{eqnarray}
where ${\tilde U}({\bf k}) = \int {\rm e}^{i {\bf k}.{\bf r} }
U({\bf r}) d{\bf r}$ is the Fourier transform of the impurity potential.
It is at this point that the three main approximations which lead to the
effective mass equation (EME) are made.
\par The first is to expand $E_{{\bf k}'}$ to second order in
${\bf k}'-{\bf k}_\mu$, around the six, degenerate, conduction band minima
located at the points
\begin{equation}
\left\{{\bf k}^{(0)}_\mu\right\}_{\mu=1..6} = \frac{2 \pi}{d}\{(0,0,\pm 0.85),(0,\pm
0.85, 0), (\pm 0.85,0,0)\},
\end{equation}
 where $d=5.43$\AA \ is the lattice constant of
silicon. This expansion allows the Bloch energies to be written in the form of an effective kinetic energy,
$E_{{\bf k}'} \approx \sum_\mu (k_\parallel'^2/(2 m^*_\parallel) +{\bf k}_\perp'^2/(2 m^*_\perp)), $ with non-isotropic
effective masses, $m_\perp^* \neq m_\parallel^*$. In this expression $k_\parallel'$ is the component
of ${\bf k'}-{\bf k}_\mu$ which is parallel to the displacement from the
origin of the $\mu^{\rm th}$ conduction band minimum in reciprocal space, ${\bf k}_\perp'$ denotes the components
perpendicular to this direction. The $m_\perp^*,m_\parallel^*$ are
the effective electron masses, which are non-isotropic, reflecting the non-isotropic nature
of the conduction band minima. This approximation holds as long as
the donor electron wave-function is sufficiently localised  around the conduction band minima, which is a
condition that seems to be well satisfied for phosphorus donors in
silicon, and yields from Eq \ref{equation:schroedinger_exact}:
\begin{equation}
E F({\bf k}') =   \sum_\mu (k_\parallel'^2/(2 m^*_\parallel) +{\bf k}_\perp'^2/(2 m^*_\perp))F({\bf k}')
+ \int \sum_{{\bf G},{\bf G}'}
A^*_{{\bf k}',{\bf G}'} A_{{\bf k},{\bf G}}
{\tilde U}({\bf k}-{\bf k}'+{\bf G}-{\bf G}')  F({\bf k}) d{\bf k}.
\label{equation:schroedinger_1}
\end{equation}

\par The second approximation is to ignore terms in potential for which
${\bf G} \neq {\bf G}'$, the assumption being that $|{\tilde U}({\bf
k}-{\bf k}'+{\bf G}-{\bf G}')| << |{\tilde U}({\bf k}-{\bf k}')|$,
over the range in which $F({\bf k})$ is significant. This is well satisfied for a Coulombic impurity potential, ${\tilde U}({\bf k}) \sim 1/{\bf k}^2$, in silicon, where the magnitude of the reciprocal lattice
vectors is  $|{\bf G}| = n 2 \pi/d$. It is, however, well known that this approximation is not consistent with the lifting of the six-fold ground-state degeneracy observed experimentally. Such a spectrum can only be produced by a potential that introduces significant valley-orbit coupling, that is, coupling between Bloch functions located at different conduction band minima, which are separated in reciprocal space by $\Delta {\bf k} \approx 1.2 \times 2 \pi/d \  {\rm and} \ 1.7 \times 2 \pi/d$, for valleys on orthogonal and parallel axes respectively. Thus any potential that is sufficiently broad in reciprocal space to produce the required valley-orbit coupling to correctly predict the low energy donor spectrum, will not satisfy this approximation in the effective-mass formalism. None the less, this approximation is inherent in the effective-mass approach and allows Eq \ref{equation:schroedinger_1} to be reduced to
\begin{equation}
E F({\bf k}') =   \sum_\mu (k_\parallel'^2/(2 m^*_\parallel) +{\bf k}_\perp'^2/(2 m^*_\perp))F({\bf k}')
+ \int \sum_{\bf G}
A^*_{{\bf k}',{\bf G}} A_{{\bf k},{\bf G}}
{\tilde U}({\bf k}-{\bf k}')  F({\bf k}) d{\bf k}.
\label{equation:schroedinger_2}
\end{equation}
\par Finally, in deriving the EME, it is assumed that the Bloch coefficients are independent of {\bf k}, $A_{{\bf k}',{\bf G}} \approx A_{{\bf k},{\bf G}}$, in the vicinity of the conduction band minima, over which the magnitude of the envelope function is significant. This is true in the immediate vicinity of the conduction band minima, but breaks down in the vicinity of the Brillouin zone boundary. This approximation, along with the identity $\sum_{\bf
G} |A_{{\bf k},{\bf G}}|^2 =1$ finally yields of
Eq~\ref{equation:schroedinger_2} the multi-valley effective-mass
equation (MV-EME):
\begin{equation}
E \sum_\mu F_\mu({\bf k}') = \hbar^2(k_\parallel'^2/(2 m^*_\parallel) +
{\bf k}_\perp'^2/(2 m^*_\perp)) \sum_\mu F_\mu({\bf k}') +
\int {\tilde U}({\bf k}-{\bf k}') \sum_\mu F_\mu({\bf k}) d{\bf k}.
\label{eq:mveme}
\end{equation}
Here we have written the envelope function, $F({\bf k}) =
\sum_\mu F_\mu({\bf k})$,  which is consistent with the final approximation above, where it assumed that the envelope functions are strongly localised around each of the conduction band minima.
\par In the absence of so-called valley-orbit coupling, that is for
potentials for which the first approximation is well satisfied, Eq~\ref{eq:mveme} decouples into
six independent single-valley effective-mass equations (SV-EME);
\begin{equation}
E  F_\mu({\bf k}') = \hbar^2(k_\parallel'^2/(2 m^*_\parallel) +
{\bf k}_\perp'^2/(2 m^*_\perp))  F_\mu({\bf k}') +
\int {\tilde U}({\bf k}-{\bf k}')  F_\mu({\bf k}) d{\bf k}.
\end{equation}
For a Coulombic impurity potential, this is isomorphic to a
non-isotropic, hydrogenic, Schr\"odinger equation in momentum space.
Such an equation cannot, in general, be solved analytically, however, for the case
of phosphorus donors Kohn and Luttinger \cite{Kohn55} proposed
solutions of the form
\begin{equation}
F_{\pm z}({\bf r}) = \frac{ \exp \left[ -\sqrt{(x^2+y^2)/a_{\perp}
+ z^2/a_{\parallel} } \,\,\right]}{\sqrt{6 \pi a_{\perp}^2
a_{\parallel}}},
\end{equation}
where $F_\mu({\bf r}) = \int {\rm e}^{i({\bf k}-{\bf k}_\mu).{\bf
r}} F_\mu({\bf k}) d{\bf r}$. The non-isotropic
effective Bohr radii, $a_\perp,a_\parallel$, are variational parameters, determined by minimising the energy of the state. 
\par The ground state donor electron wave-function, found in this way, is six-fold degenerate, and is given by linearly independent combinations of the functions
\begin{equation}
\psi({\bf r})_\mu = F_\mu({\bf r}) \phi_{{\bf k}_\mu}({\bf r}).
\end{equation}
As mentioned earlier, this six-fold degeneracy of the ground state is at odds with
experimental observation that the ground state is a singlet with
binding energy $45.5$ meV, lying $11.85$ meV below a triplet state
which is, in turn, $1.42$ meV below a doublet state. This lifting of
the ground-state degeneracy can be predicted from group-theoretical
considerations, due to the breaking of the crystal symmetry. The splitting
can only be produced by a  potential that is strong enough to couple the
different valleys. Group theoretical arguments give the low-lying energy
states as
\begin{equation}
\psi({\bf r})^i = \sum_\mu \alpha^i_\mu F_\mu({\bf r}) \phi_{{\bf k}_\mu}({\bf r}),
\end{equation}
where the coefficients are given by

\begin{equation}
  \begin{array}{lll}
    \begin{array}{l}
       {\boldsymbol \alpha}^1 = \frac{1}{\sqrt{6}}(1,1,1,1,1,1) \\
    \end{array}
    \left.\begin{array}{l}\ \\ \end{array}\right\} &
    A_1 \\[20pt]
    \begin{array}{l}
      {\boldsymbol \alpha}^2 =    \frac{1}{\sqrt{12}}(-1,-1,-1,-1,2,2) \\
       {\boldsymbol \alpha}^3 =    \frac{1}{2}(1,1,-1,-1,0,0) \\
    \end{array} &
    \left.\begin{array}{l}\ \\ \ \\ \end{array}\right\} &
    E \\[20pt]
    \begin{array}{l}
       {\boldsymbol \alpha}^4 =  \frac{1}{\sqrt{2}}(1,-1,0,0,0,0) \\
       {\boldsymbol \alpha}^5 =  \frac{1}{\sqrt{2}}(0,0,1,-1,0,0) \\
       {\boldsymbol \alpha}^6 =  \frac{1}{\sqrt{2}}(0,0,0,0,1,-1) \\
    \end{array} &
    \left.\begin{array}{l}\ \\ \ \\ \ \\ \end{array}\right\} &
    T_1 \\
  \end{array}
\label{eq:alpha}
\end{equation}
Here the labels on the right denote the irreducible representation of
the $T_d$ symmetry group to which the states belong, states in the
same representation are degenerate. This wave-function is the
cornerstone of the effective mass formalism, and it is axiomatic
within the theory that the ground-state wave-function is of this
form, however, the effect of the valley-orbit coupling has been handled in a rather {\it ad hoc} manner. 
While the effective-mass formalism can be used to give a reasonable prediction of the low-lying energy spectrum, with the inclusion of 
a phenomenological valley-orbit term in the potential, the inclusion of such a term is inconsistent with the approximations inherent in the effective-mass procedure.

\section{\label{sec:bloch} Beyond the effective-mass formalism}
To correctly predict the low-lying energy spectrum of a phosphorus donor electron in silicon, it is necessary to
treat the inter-valley, or valley-orbit, coupling, produced by the impurity potential. In the previous section we have argued that the treatment of such a potential is beyond the scope of some of the approximations that go into deriving the EME. Therefore, instead of looking for solutions of the EME, we take the approach of directly solving the exact Schr\"odinger equation for the donor electron, Eq \ref{equation:schroedinger_exact}. This is done by discretising the integral equation
to reduce the problem to a that of finding the eigen-solution of the Hamiltonian matrix in a truncated basis of Bloch states. Our method proceeds as follows: The discretised version of the Schr\"odinger equation, Eq(\ref{equation:schroedinger_exact}), can be written as a matrix equation;
\begin{equation}
E f_{{\bf k}'} = E_{{\bf k}'} f_{{\bf k}'} +  \sum_{{\bf k},{\bf G},{\bf G}'}
\sqrt{\omega_{\bf k} \omega_{{\bf k}'}}   A^*_{{\bf k}',{\bf G}'} A_{{\bf k},{\bf G}}
{\tilde U}({\bf k}-{\bf k}'+{\bf G}-{\bf G}')  f_{\bf k} ,
\end{equation}
where $\omega_{\bf k} \propto \Delta {\bf k}$ is an integration constant, and  $f_{\bf k} = F({\bf k})\sqrt{\omega_{\bf k}}$. Of course, in the limit that we sample ${\bf k}$ at an infinitely
dense set of points along the entire conduction band, this equation
is completely equivalent to Eq(\ref{equation:schroedinger_exact}).
Computational limitations restrict us to much more limited
set of sample points, which need to be chosen carefully to optimise the accuracy of the solution. 
\par We initially solve the problem for the
Coulombic impurity potential, before proceeding to include
corrections. The use of a Coulombic potential $U({\bf q}) =
2/(\kappa \pi^2  {\bf q}^2)$, presents a difficulty, namely the
presence of a singularity at ${\bf q}=0$. This singularity is, of
course, integrable, however, it does present problems for the
discretised equation. The singularity can be treated by the
use of an appropriate regularising function \cite{Krautgartner92}
\begin{equation}
S({\bf q},{\bf q}') = \frac{(1/ \alpha^2 + {\bf q}^2)^2}{(1/ \alpha^2 + {\bf q}'^2)^2}.
\end{equation}
This function has the key properties that $S({\bf q},{\bf q})=1$, and  it is analytically integrable
\begin{equation}
\int S({\bf q},{\bf q}')/({\bf q} - {\bf q}')^2 d{\bf q}' = \pi^2 (1/ \alpha + \alpha q^2).
\end{equation} 
The singularity is removed by addition and subtraction of this regularising function into the summand. If we ignore the possibility that ${\bf k}-{\bf k}'+{\bf G}-{\bf G}' =0 , {\bf G} \neq {\bf G}'$, which is a weaker form of approximation two in the effective-mass formalism above, this gives;
\begin{eqnarray}
E f_{{\bf k}'} = E_{{\bf k}'} f_{{\bf k}'} &-& \frac{1}{\kappa \pi^2} \sum_{\bf k}
\Big \{ \sum_{{\bf G} \neq {\bf G}'}   \frac{ \sqrt{\omega_{\bf k} \omega_{{\bf k}'}} A^*_{{\bf k}',{\bf G}'} A_{{\bf k},{\bf G}}}{({\bf k} -{\bf k}'+{\bf G} -{\bf G'})^2 } f_{\bf k} \nonumber\\
&-&   \frac{   A^*_{{\bf k}',{\bf G}} A_{{\bf k'}_{\bf G}}
 (f_{\bf k} \sqrt{\omega_{\bf k} \omega_{{\bf k}'}} -   f_{{\bf k}'} \omega_{\bf k} S({\bf k}',{\bf k}))} {({\bf k} -{\bf k}')^2} \Big \} \nonumber\\
&-& \frac {f_{{\bf k'}}  (1/ \alpha + \alpha k')}{\kappa}.
\end{eqnarray}
 The parameter $\alpha$, in the regularising function, can be adjusted to optimise the efficiency of the method. In the case of an isotropic hydrogenic Schr\"odinger equation, setting $\alpha = \kappa/m^*$,  where $\kappa$ is the dielectric constant of the material and $m^*$ is the (isotropic) effective mass, is optimal. In our case, we use $m^* = m_\perp = 0.191$.

The set of points chosen, in the case of the bare Coulombic
potential, is guided by the optimised Kohn-Luttinger solution to the
SV-EME, which we expect to be a good approximation  for the exact solution using this potential. Accordingly, we choose points located around the six
conduction band minima. Due to the known non-isotropy of the EME
solution, we find that best results are obtained when the spacing of
points in the longitudinal direction around each minimum are 1.7
times greater than those in the transverse directions.
The points are chosen in such a way that $\Delta
k_\parallel = 0.29 \times 2 \pi/(d(N-1)) , \Delta k_\perp = \Delta k_\parallel/1.7$, with the total number of points $N_t =
6\times N^3$. Points chosen in this way ensure that a point is taken
at each of the conduction band minima, and that a point is taken close to  the edge of the Brillouin zone. In practice, memory
limitations restrict $N\leq 11$. In Table(\ref{tab:energies1}) we
present ground state binding energies obtained using both the
non-isotropic sampling, and the isotropic ($\Delta k_\perp = \Delta
k_\parallel$). The non-isotropic sampling produces the best results
in this situation, and for $N=11$ the results seem to have converged, and are
in good agreement with those obtained from effective mass theory.

\begin{table}[h]
\caption{\label{tab:energies1} Ground state binding energies (meV)
obtained for single and multi-valley methods using isotropic and
non-isotropic point sampling as described in the text. Compare these
with 28.9 meV for the multi-valley EME and for the single-valley
EME.}
\begin{ruledtabular}
\begin{tabular}{cccccccc}
 N & 5 & 7 & 9 & 11 & 13 & 15 & 17\\
\hline
MV non-isotropic & 19.6 & 25.0 & 29.3 & 30.3 &- &- &-\\
MV isotropic  & 19.3 &  20.6 & 25.4 &28.6 &- &- &-\\
SV non-isotropic & 19.6 & 24.9 & 29.1 & 30.1 &30.3 &30.4 &30.4\\
SV isotropic  & 19.3 &  20.6 & 25.3 &28.4 &29.5 &29.8 &29.9\\
\end{tabular}
\end{ruledtabular}
\end{table}

\section{\label{sec:correction} Corrected impurity potentials}

Pantelides \cite{Pantelides74} introduced a  non-Coulombic correction to the impurity potential based on the non-static properties of the dielectric function:
\begin{equation}
\frac{1}{\epsilon({\bf q})} = \frac{A q^2}{q^2+\alpha^2} + \frac{(1-A) q^2}{q^2+\beta^2} + \frac{1}{\epsilon(0)} \frac{ \gamma^2}{q^2+\gamma^2},
\end{equation}
where $\epsilon(0) = 11.9 \epsilon_0$ is the static dielectric
constant for silicon, with the parameters,  in atomic
units, $A=1.175, \alpha = 0.7572, \beta = 0.3123,$ and $\gamma =
2.044$. If considered as a correction to a Coulombic potential with
static dielectric constant, the corrected potential can be written
\begin{equation}
U_{\rm cor}({\bf q}) = \frac{1}{\pi^2 \kappa} \left(\frac{\kappa A q^2}{q^2+\alpha^2} + \frac{\kappa (1-A) q^2}{q^2+\beta^2} - \frac{ q^2}{q^2+\gamma^2} \right).
\end{equation}
In our work we take the liberty of varying the strength of this
correction, using a parameter $\eta$, which we vary to obtain a
ground state binding energy in agreement with experiment. The
corrected potential has a broader Fourier spectrum and introduces
coupling between different valleys, breaking the ground state
degeneracy that the Coulomb potential fails to significantly lift.
It is also found that the solutions the Schr\"odinger equation for
the corrected potential are more diffuse in the reciprocal space,
that is they extend over greater values of ${\bf k}$, and are more
isotropic. This encourages us to use the isotropic sampling
procedure to ensure that points are taken over a sufficient range to
cover the wave-function solution. A correction
strength of $\eta = 5.8$ is found to give a ground state binding
energy in agreement with that observed experimentally, with the
excited state energies also in relatively good agreement, as seen in Table(\ref{tab:energies2}), and the correct multiplicities predicted. 

\begin{figure}
\rotatebox{-90}{\resizebox{6cm}{!}{\includegraphics{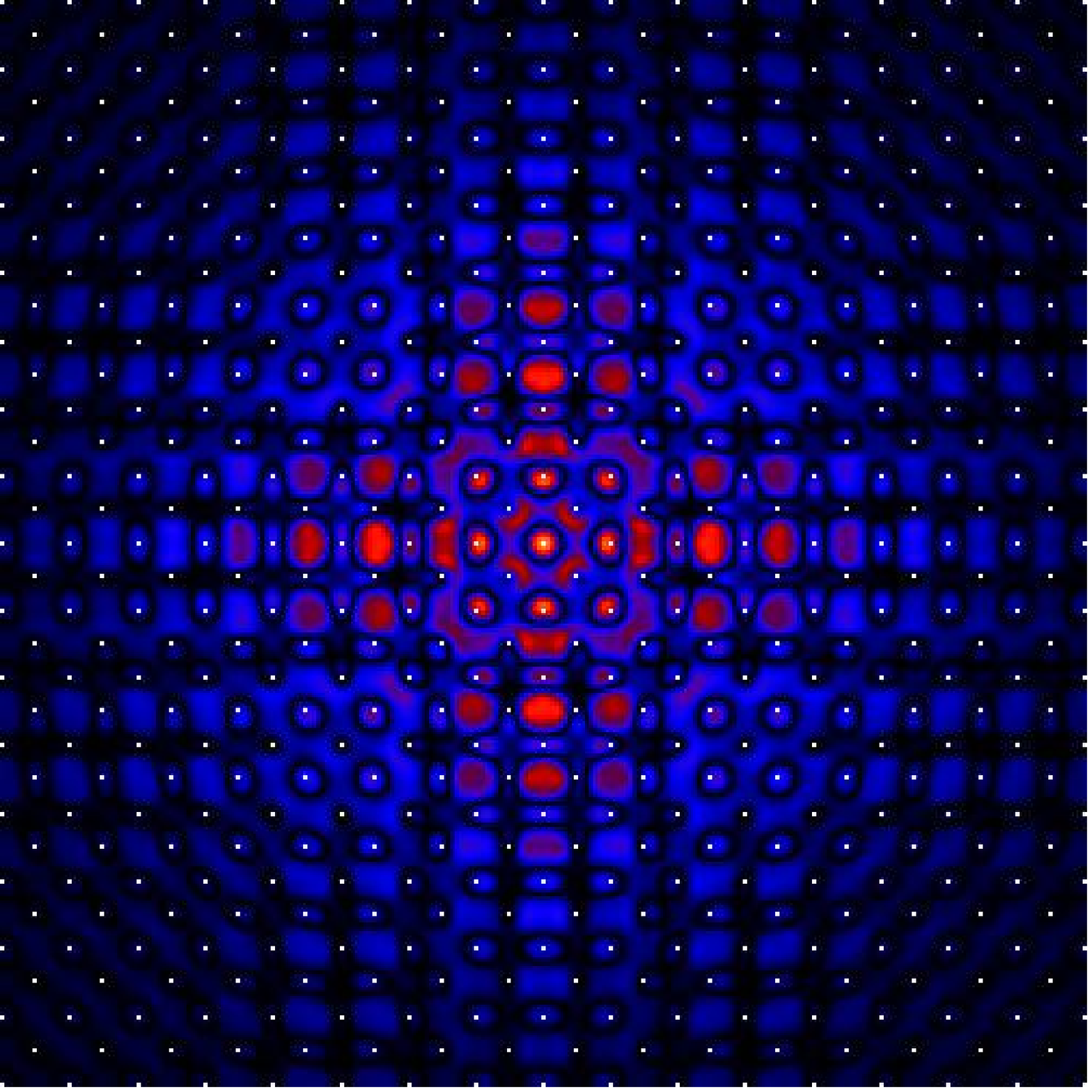}}}
\rotatebox{-90}{\resizebox{6cm}{!}{\includegraphics{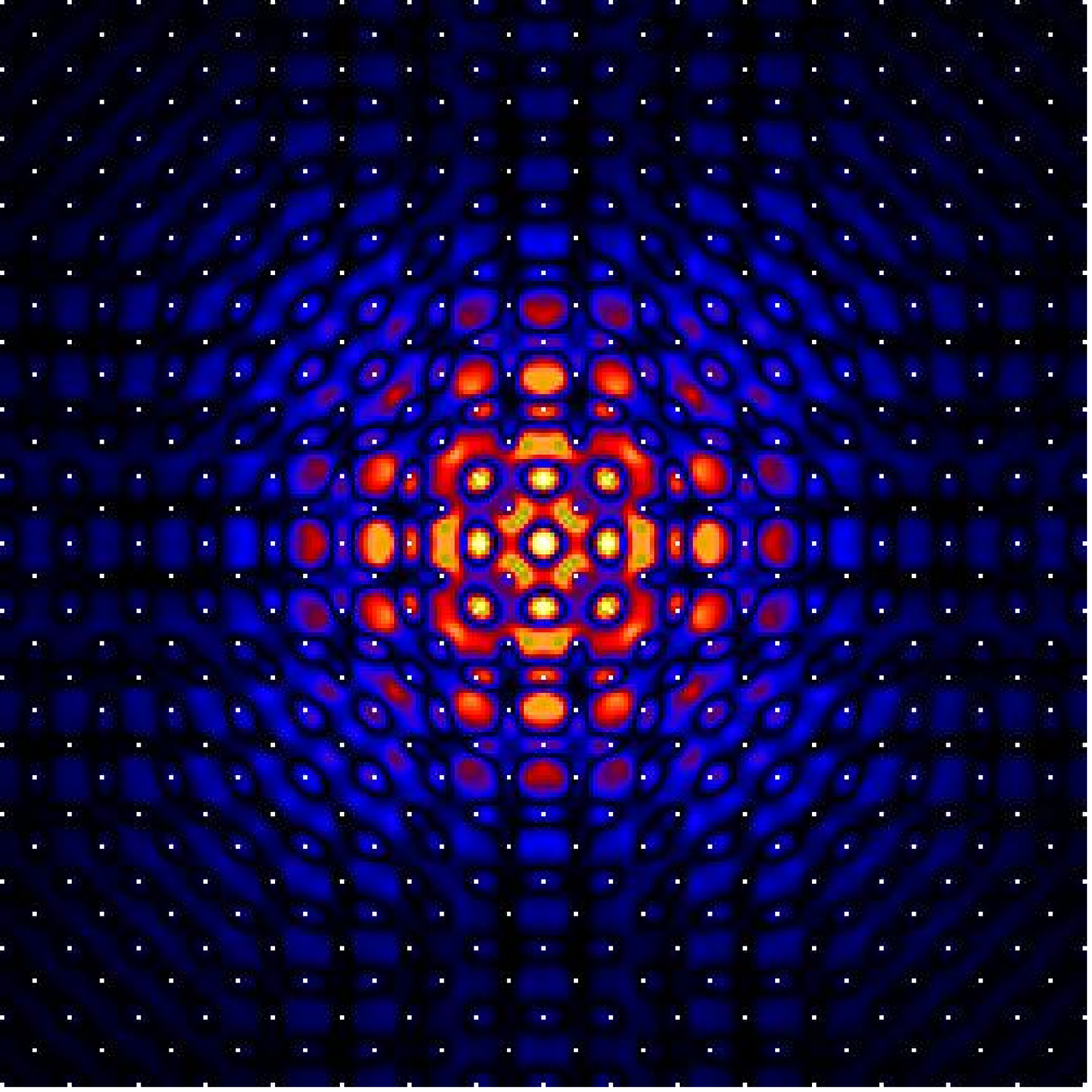}}}
\caption{(Colour online) Modulus of the donor electron wave-function in the $[001]$
plane for the case $N=11$, with $\eta =0$, and $\eta = 5.8$. The
white dots mark atomic sites of the silicon lattice.} \label{fig:WF}
\end{figure}

We point out here that similar corrections have been made within the effective-mass approximation, for examples see \cite{Ning71,Pantelides74}, with equally good agreement obtained for the low-lying energy states. However, as discussed previously, such an approach is not consistent with the approximations inherent in the effective-mass approach. The numerical approach outlined in this article, however, allows the calculation of the donor wave-function for impurity potential of any form. In Fig \ref{fig:WF} we show the modulus of the wave-function in real space, obtained from numerical solution of the donor Hamiltonian, both for the Coulomb potential, and the corrected, $\eta = 5.8$, potential. Here the effect of the corrected potential is obvious, in that it more strongly localises the donor state around the donor impurity, consistent with what we would expect from a more tightly bound state.

\begin{table}
\caption{\label{tab:energies2}
Comparison between energies calculated using $N = 11$ and experimentally observed values. Energies are quoted in meV.}
\begin{ruledtabular}
\begin{tabular}{cccc}
&1S(A)&1S(T)&1S(E)\\
\hline
Theory ($\eta = 5.8$) & 45.5 & 29.1 & 27.1 \\
Experiment \footnote{experimental data ref. \cite{Aggarwal65} corrected using 3p$_{\pm}$ state from ref. \cite{Faulkner69} }& 45.5 & 33.6 & 32.2 \\
\end{tabular}
\end{ruledtabular}
\end{table}

\section{\label{sec:exchange}exchange}

The increase of the localisation of the electron wave-function around the donor nucleus, which is caused by the added correction to the donor potential necessary to lift the ground-state degeneracy, will have consequences on the construction of a spin-based phosphorus donor quantum computer. The exchange energy between neighbouring donor electrons, which mediates the inter-qubit interactions in most Si:P QC proposals, depends on the overlap of the two electronic wave-functions. It is, therefore, necessary to calculate the exchange coupling between these donors, using wave-functions have the correct ground-state energy. This has not, to the knowledge of the authors, been done until now, and cannot be easily done in the effective-mass formalism where the effect of the valley-orbit coupling on the donor wave-functions is not treated consistently. 

\begin{figure}
\rotatebox{-90}{\resizebox{6cm}{!}{\includegraphics{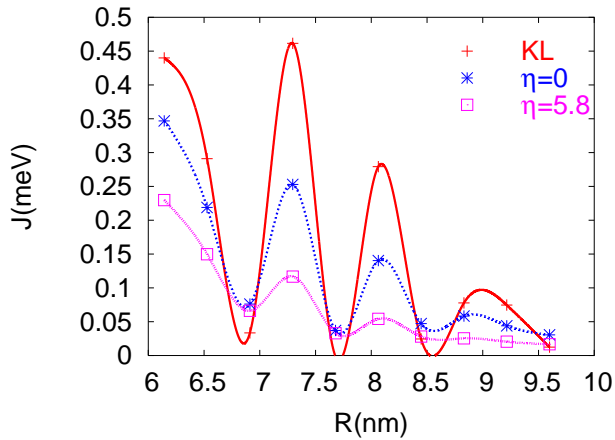}}}
\caption{(Colour online) The strength of the exchange coupling between neighbouring donors located in the same [001] plane as a function of their separation along the crystallographic [110] axis. The plot compares results calculated using the Kohn-Luttinger effective-mass wave-function, as well as wave-functions obtained using the numerical technique described in this article, both with and without a corrected impurity potential. Note that the points refer to substitutional sites in the silicon matrix, the lines between these points are a guide to the eye only, the wave-functions used are only valid for substitutional donors.}
\label{fig:exchange_eta}
\end{figure}

To this end, we calculate the exchange coupling between neighbouring donor electrons, using the Heitler-London approach, as a function of the donor separation in the $[110]$ direction Fig \ref{fig:exchange_eta}, using the wave-functions calculated in the manner described previously. We find, as expected, that
the corrected potential reduces the strength of this coupling, due to the increased localisation of the electron states. 
We compare our results to those obtained using solutions of the SV-EME, and find that the strength of the exchange coupling is, in general, reduced. It is also pertinent to note that although still present, the oscillatory behaviour of the exchange coupling with donor separation is somewhat reduced in amplitude when calculated using states obtained by solution of the exact Hamiltonian. This expected, and comes from the fact that the wave-function is not as strongly localised around the conduction-band minima as was the case for the effective-mass solution, thus reducing the strength of the interference.

\section{\label{sec:strain}Lattice strain}
It has been pointed out that the oscillatory nature of the spatial dependence of the exchange coupling between donors can be partially suppressed by the application of strain to the silicon substrate \cite{Koiller02B,Keyes02}. Koiller {\it et al} \cite{Koiller02B} have calculated the exchange splitting for donors located in the same [001] plane, parallel to the direction of a tensile strain produced by over-growth of the silicon substrate on a relaxed Si$_{0.8}$Ge$_{0.2}$ substrate. This uni-axial strain breaks the tetrahedral symmetry of the silicon crystal, and thus lowers the conduction band energy minima in the direction perpendicular to the over-growth, relative to those that are parallel. This ensures that the ground-state wave-function has contributions from the two perpendicular minima alone, eliminating the oscillatory dependence of the exchange coupling for displacements within the plane. 

\begin{figure}
\rotatebox{-90}{\resizebox{6cm}{!}{\includegraphics{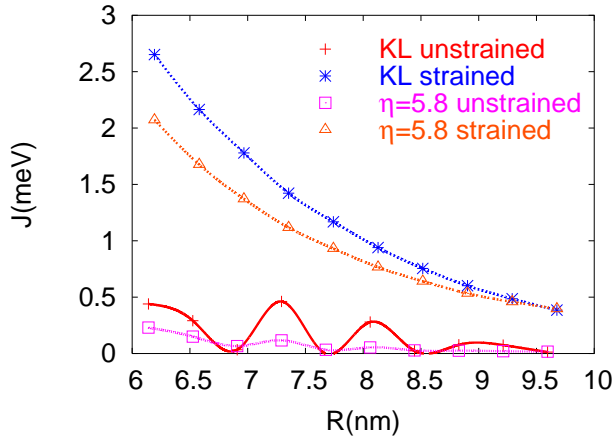}}}
\caption{(Colour online) The strength of the exchange coupling between neighbouring donors located in the same [001] plane as a function of their separation along the crystallographic [110] axis. The plot compares results for strained and unstrained silicon, calculated using both the effective-mass theory and the wave-functions obtained as described in this paper. The tensile strain is obtained by over-growth on a relaxed [001] Si$_{0.8}$Ge$_{0.2}$ surface. Note that the points refer to substitutional sites in the silicon matrix, the lines between these points are a guide to the eye only and do not have any physical meaning, as the wave-functions are only valid for substitutional donors}
\label{fig:exchange_strain}
\end{figure}

These effects can easily be incorporated into our calculation simply by re-calculating the silicon band-structure in the presence of strain, which we achieve by altering the silicon lattice spacings in the different directions  appropriately. The donor electron wave-function is then expanded in the basis of these strained Bloch functions, and used to calculate the exchange coupling in the Heitler-London approximation. To illustrate this we have reproduced some of the results of Koiller and calculated the exchange coupling as a function of separation along the [110] axis parallel to the plane of the tensile strain. The results are shown in Fig \ref{fig:exchange_strain} where it can be seen that the exchange coupling strength decays monotonically with the separation distance, as opposed to the unstrained case in which oscillatory behaviour is observed. In agreement with previous results \cite{Koiller02B,Koiller04}, we see that the strain removes the oscillatory dependence of the exchange coupling on the in-plane donor separation, of course the dependence will still be oscillatory for displacements out of the plane. We see also, in agreement with our previous discussion, that the exchange coupling calculated for the core-corrected potential is slightly less than that predicted by the effective-mass theory.

\section{\label{sec:epot}applied electro-static potentials}
An important problem in the construction of a phosphorus-in-silicon quantum computer is to calculate the response of the donor electron wave-function to electro-static potentials created by voltage biases applied to control gates. It is through the application of such biases that coherent control of the electron wave-function is to be achieved and quantum information processing implemented.  In both nuclear-spin \cite{Kane98}, and some electron-spin \cite{Hill04} proposals, one of the key issues is the control of the electron-nuclear contact-hyperfine coupling strength with externally applied voltage biases. The dependence of the hyperfine coupling strength on the applied bias has been studied, using  simplified hydrogenic type donor wave-functions in the presence of realistic control-gate biases \cite{Wellard02,Kettle03}, as well as for more sophisticated tight-binding wave-functions in the presence of simple uniform electric fields \cite{Martins04}.

We illustrate the flexibility of this technique of direct numerical diagonalisation of the system Hamiltonian in the basis of Bloch states by using our technique to reproduce the results of Martins {\it et. al.} \cite{Martins04}. This is easily achieved by simply including the potential term due to the uniform electric field in the Hamiltonian to be diagonalised. As was the case in the tight-binding approach, a slight complication arising from the choice of boundary conditions has an accidental realistic physical interpretation. In our case we assume that the electric field is being applied in the positive $z$ direction. To make the calculation tractable we must assume that the field is finite in extent, this is obviously physically reasonable, and we vary extent of the field. This leads to a discontinuity in the potential at the edge of this region. At the $-z$ edge this discontinuity qualitatively reflects the barrier potential an electron would experience due to the presence of a silicon-oxide barrier, as would be present in a MOS device.

\begin{figure}
\rotatebox{-90}{\resizebox{6cm}{!}{\includegraphics{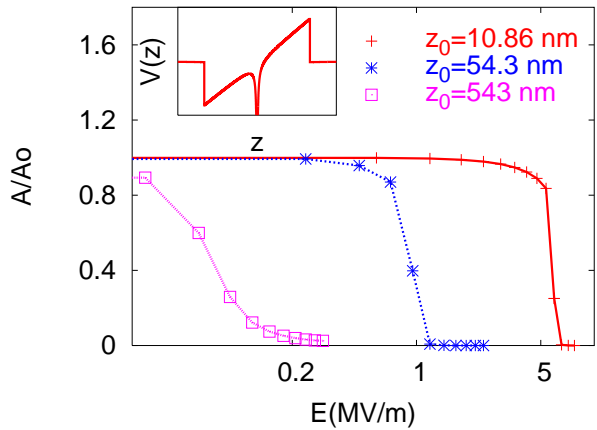}}}
\caption{(Colour online) The field dependence of the electron-nuclear contact hyperfine coefficient (A), in units of the zero-field coupling (A$_0$), for donors located at different distances from an oxide barrier (Z$_0$), the lines between points are a guide to the eye. Donors closer to the barrier require a stronger field to significantly deform the wave-function} 
\label{fig:A}
\end{figure}

The contact-hyperfine coupling strength between the donor and the nucleus is proportional to the probability of the electron being found at the position of the nucleus, thus to calculate the change in the coupling constant as a function of applied field, it is necessary to calculate the electron wave-function for various field strengths. This has been done using the corrected Coulomb potential, $\eta = 5.8$, for donors located at various distances from the silicon-oxide barrier. We show our results, normalised with respect to the zero-field case, in Fig.~\ref{fig:A}. The fact that the pseudo-potential approach used to calculate the Bloch structure of the silicon lattice is insufficiently sophisticated to give a good description of the electron wave-functions in the vicinity of the silicon ion cores means that the absolute values calculated for the hyperfine interaction are not in agreement with experimental observation, and only relative shifts of this quantity can be calculated with any confidence. The results are in good agreement with those obtained by Martins {\it et. al.} in a tight-binding study, and indicate that donors close to the silicon-oxide barrier require a greater field strength to achieve a certain shift in the hyperfine coupling, than do those that are further away. This is simply due to the intuitive fact that the potential well created by the applied field at the oxide boundary, is relatively shallower for donors close to the barrier, than it is for those further from the barrier.

\section{Conclusions and outlook}
We present a method for calculating the electronic states of phosphorus donors in silicon by numerical diagonalisation of the exact donor Hamiltonian, in a truncated basis of Bloch states. This technique avoids the approximations inherent in the effective-mass formalism and allows the calculation of eigen-states for a variety of potentials, including corrected impurity potentials, as well as applied electro-static potentials. We use a correction impurity potential to calculate the low-lying energy spectrum of the phosphorus donor electron and obtain good agreement with experiment. We then use the wave-functions obtained in this manner to calculate the exchange coupling strength between neighbouring donor electrons, and find that the valley-orbit coupling reduces the strength of this interaction due the increased localisation of the states. The effects of lattice strain are incorporated in the calculation and the exchange coupling between donors in a strained substrate is calculated. The ability of the technique to include the effects of externally applied electro-static potentials is illustrated by calculating the electron-nuclear hyperfine coupling constant as a function of applied uniform field.
\par  Ultimately the usefulness of this method is determined by the number of points in reciprocal space that can be included in the Hamiltonian matrix. We have managed to include up to 7986 points. While this seems to be enough to obtain convergence for the energy, the range over which the real-space wave function can be reproduced with this number of points is limited, by aliasing, to a distance of approximately 150 \AA \ from the position of the donor nucleus. To calculate exchange couplings for donors separated by distances of more than this will require the inclusion of more points in the Hamiltonian.
\par Finally we wish to emphasise that the results obtained using this technique vary only quantitatively from what is predicted using effective-mass theory. These quantitative differences are, however, important for quantum information applications, where the demands on precision are high.

\section{Acknowledgements} 
This work was supported by the Australian Research Council, and the Army Research Office under contract DAAD19-01-1-0653. Computational support was provided by the Australian Partnership for Advanced Computing, as well as the Victorian Partnership for Advanced Computing. The authours would also like to acknowledge valuable discussions with Hsi-Sheng Goan and Louise Kettle.


\end{document}